\documentclass[9pt,letterpaper,twocolumn]{article}
\usepackage{amssymb}
\usepackage{sectsty}
\usepackage{amsmath}
\usepackage{epsfig}
\usepackage{ifthen}
\usepackage{law}
\usepackage{graphics}
\usepackage{times}
\usepackage{cite}
\usepackage{mathptmx}
\sectionfont{\centering\small}
\subsectionfont{\small}

\newcommand{\avg}[1]{\left\langle{#1}\right\rangle}

\newcommand{\be}{\begin{equation}}
\newcommand{\ee}{\end{equation}}
\newcommand{\bed}{\begin{displaymath}}
\newcommand{\eed}{\end{displaymath}}
\newcommand{\beas}{\begin{eqnarray*}}
\newcommand{\eeas}{\end{eqnarray*}}
\newcommand{\bea}{\begin{eqnarray}}
\newcommand{\eea}{\end{eqnarray}}

\newcommand{\comment}[1]{}

\title{
B-Rank: A top $N$ Recommendation Algorithm}

\author{ 
Marcel Blattner\\
Laboratory for Web Science, University of Applied Science FFHS\\ 
Regensdorf, Switzerland}

\date{}
\begin{document}
\maketitle
\thispagestyle{empty}
\section*{ABSTRACT}
In this paper B-Rank, an efficient ranking algorithm for recommender systems, is proposed. B-Rank is based on a random walk model on hypergraphs. Depending on the setup, B-Rank outperforms other state of the art algorithms  in terms of precision, recall $\sim (19\% - 50\%)$ and inter list diversity $\sim (20\% - 60\%)$. B-Rank captures well the difference between popular and niche objects. The proposed algorithm produces very promising results for sparse and dense voting matrices.  Furthermore, a recommendation list update algorithm is introduced,to cope with new votes. This technique significantly reduces computational complexity. The algorithm implementation is simple, since B-Rank needs no parameter tuning.
\section{INTRODUCTION}
One of the most amazing trends of today's globalized economy is
{\em  peer production}~\cite{An06}.
An unprecedented mass of unpaid workers is
contributing to the growth of the World Wide Web: some build
entire pages, some only drop casual comments, having no other
reward than {\em reputation}~\cite{MaZh04}.
Many successful web sites (e.g. Blogger and MySpace) are just platforms
holding user-generated content.
The information thus conveyed is particularly valuable because it
contains personal opinions, with no specific corporate interest. It is,
at the same time, very hard to go through it and judge its degree of
reliability. If you want to use it, you need to filter this
information, select what is relevant and aggregate it; you need to
reduce the information overload~\cite{Ma94}.

As a matter of fact, opinion filtering has become rather common on the
web. There exist search engines (e.g. Google news) that are able to extract news
from journals, web sites (e.g. Digg) that harvest them from blogs, platforms
(e.g. Epinions) that collect and aggregate votes on products. The basic
version of these systems ranks the objects once for all, assuming they
have an intrinsic value, independent of the
personal taste of the demander~\cite{LaMo06}. They lack
{\em personalization}~\cite{Ke06}, which constitutes the
new frontier of online services.

Personal information can, in fact, be
exploited by {\em recommender systems}.
Amazon.com, for instance, uses one's purchase
history to provide individual suggestions. If you have bought a
physics book, Amazon recommends you other physics books: this is
called  item-based or content-based recommendation~\cite{BrHe98,SaKa01}.
Many different techniques have been developed in the past, including
collaborative filtering methods \cite{ReIa94,BrHe98,ClGo99,GoSc99},
content-based techniques \cite{Pa99,MeMo02,SoHo99,Pa07}, spectral based
methods \cite{BeNi03,GoRo01,BlHuLa05} and network based algorithms \cite{ZhBlYk07,ZhJiZh08,ZhKuLiMe09}.

The evaluation of recommender systems is difficult \cite{HeKo04}. There are several
reasons for this: a) an algorithm may perform well on a particular
data set but fails on others, b) in the past, evaluations focused on
predictive accuracy of withheld ratings. Novelty and diversity were mostly ignored.
These two factors play a pivotal role from a user point of view \cite{NeRiKo06}.
c) there is still no common framework in the community, defining a
set of evaluation metrics. Such a framework would be particular useful, when comparing
different techniques based on different data sets.

In this paper B-Rank, a novel top $N$ recommendation algorithm, is presented. B-Rank is
based on a markov chain model \cite{kemeny1} on hypergraphs \cite{DiMa}. The algorithm produces high precision
and recall performance, maintaining high diversity between different recommendation lists at the same time.
B-Rank is parameter free, which is very attractive from an implementation point of view. The
performance is measured on two complementary data sets - movielens and jester. B-Rank is compared to collaborative filtering method and ZLZ-II \cite{ZhLuZh09,ZhKuLiMe09}, which is known to be superior to ordinary recommendation algorithms the investigated setup. GRank \cite{ZhKuLiMe09}, a global ranking method, serves as a base benchmark.

The paper is organized as follows. Sec.(\ref{sec:meth}) outlines all methods, test procedures and data set descriptions  used in the paper. Sec.(\ref{sect:results}) contains the main results: numerical performance evaluations for all methods and corresponding comparisons as well. Furthermore, results on computational complexity reduction and an efficient update algorithm are presented. Sec.(\ref{sect:disc}) contains a discussion of the results and in Sec.(\ref{sect:sum}) a summary and outline for future are given. Additional remarks and explanations on B-Rank are presented in Sec.\ref{sect:addmat}.  
\section{METHODS \label{sec:meth}}

User ratings are stored in a matrix $V(O \times U)$.
$O$ denotes the number of objects and $U$ is the number of users.
$v_{\alpha i} \in V$ corresponds to user $i$'s rating to object $\alpha$. Throughout
this paper objects are labeled by Greek letters, whereas people are identified by Latin letters.
\subsection*{B-Rank \label{subsect:brank}}
B-Rank is based on a random walk model with given initial conditions $\chi_{(i)}$, taking place on a hypergraph $\mathcal{G}$
\footnote{A hypergraph $\mathcal{G}(V,E)$ is a finite set $V$ of vertices together with a finite multiset $E$
of hyperedges, which are arbitrary subsets of $V$. The incidence matrix $H$ of a hypergraph $\mathcal{G}(V,E)$ with $E=\{e_{1},e_{2},\cdots,e_{m}\}$
and $V=\{v_{1},v_{2},\cdots,v_{m}\}$ is the $m \times m$ matrix with $h_{ij}=1$ if $v_{j} \in e_{i}$ and $0$ otherwise.} .
A transition matrix $P$ is associated with the hypergraph $\mathcal{G}$. $P_{\alpha \beta} \in P$ denotes the transition probability from
object $\alpha$ to object $\beta$.
$\chi_{(i)}$ is a normalized column vector, representing user $i$'s preference, i.e the collection of already voted objects:
$\chi_{(i)\beta} = 1/\sum_{\alpha}sign(v_{\alpha i})$, where
$sign(x)=1$ if $x>0$, and $0$ otherwise.

In the hypergraph framework, each user $i$ is modeled as a hyperedge and each object $\alpha$ is a hypergraph vertex. The transition matrix $P$ of $\mathcal{G}$ is defined like:
\begin{equation}
\label{equ:brank}
P_{\alpha \beta} = (1- \delta_{\alpha \beta})\frac{1}{k_{\alpha}}\sum_{i} w_{i}h_{\alpha i}h_{i \beta}.
\end{equation}
Where $k_{\alpha}=\sum_{\beta \ne \alpha}\sum_{i} w_{i}h_{\alpha i}h_{i \beta}$, $\delta_{\alpha \beta}$ is the Kronecker Delta and $w_{i}$ is the associated weight to hyperedge $i$.
In matrix formulation $P$ reads as $P=D^{-1}A$ with the symmetric adjacency matrix $A=HWH^{T} - T$. $D$ is a diagonal matrix containing the row sums  of
$A$, $d_{\alpha} = \sum_{\beta}A_{\alpha \beta}$. $H$ is the incidence matrix and $H^{T}$ its transposed. $W$ is the diagonal hyperedge weight matrix and $T$ is the diagonal vertex
degree matrix with $T_{\alpha}=\delta_{\alpha \beta}\sum_{i}w_{i}h_{i\alpha}h_{i \beta}$.  $P$ is a row stochastic matrix with zero diagonal by construction.

B-Rank calculates user $i$'s recommendation list $f_{(i)}$ as follows:
\begin{enumerate}
\label{step:brank}
\item Forward propagation: $f_{(i)}^{F} = P^{T}\chi_{(i)}$
\item Backward propagation: $f_{(i)}^{B} = P\chi_{(i)}$
\item Final ranking: $f_{(i)} = f_{(i)}^{F}\#f_{(i)}^{B}$ \footnote{\#denotes the element-wise multiplication of two vectors.}
\item Set already voted objects to zero.
\item Sort $f_{(i)}$ in descending order.
\item Select top $N$ items of the sorted list $f_{(i)}$
\end{enumerate}
In this paper the unweighted version of B-Rank is investigated. Therefore, the weights are set to one -$w_{i}=1$ $\forall i$. This is done, to have straight forward comparisons
to similar algorithms. 
\subsection*{ZLZ-II}
 ZLZ-II \cite{ZhKuLiMe09,ZhLuZh09}, is based on a lazy random walk process, taking place on a bipartite user-object graph. ZLZ-II uses a coarse grained version
of the original voting matrix. $a_{\alpha i} = 1$ if $v_{\alpha i} > v_{tr}$, and $0$ else. $v_{\alpha i}$ is the original vote
and $a_{\alpha i}$ is the transformed vote used in ZLZ-II.
$v_{tr}$ is a threshold, to be selected. In general only ``positive'' votes are kept and the rest is discarded. Objects are assigned with a initial ``resource'' $f$. The given resources $f$ are re-distributed according the linear transformation: $\bar{f}_{i}= Wf_{(i)}$. The resulting  $\bar{f}_{i}$ is user $i$'s
recommendation list. Like in B-Rank, this vector is sorted in descending order and the top $N$ objects are presented as recommendations.
$W$ is a column stochastic matrix:
\begin{displaymath}
W_{\alpha \beta} = \frac{1}{k_{\beta}}\sum_{j}\frac{a_{\alpha j}a_{\beta j}}{k_{j}}.
\end{displaymath}
$k_{\beta}$ is the number of votes given to object $\beta$ and $k_{j}$ is the number of objects voted by user $j$. Note, that $W$'s diagonal is non-zero (lazy random walk).
\subsection*{Collaborative Filtering}
Collaborative Filtering is perhaps the most popular recommendation method \cite{AdTu05}. It is based on user-user linear correlations as a similarity measure. 
\begin{displaymath}
v_{j \beta}' = \avg{v_j} + \sum_{i=1}^N S_{j i} (v_{i \beta}-\avg{v_i}),
\end{displaymath}
where $v_{j \beta}'$ is the predicted vote, $\avg{v_i}$ the average vote expressed by user $i$ and $S$ is the similarity matrix. A common correlation measure (pearson correlation \cite{NumRec}) is used, to calculate $S$.
\begin{displaymath}
S_{i j} = {{\sum_{\alpha}  (v_{i \alpha}-\avg{v_i}) (v_{j
    \alpha}-\avg{v_j}) } \over {\sqrt{\sum_{\alpha}  (v_{i
      \alpha}-\avg{v_i})^2} \sqrt{\sum_{\alpha}  (v_{j
      \alpha}-\avg{v_j})^2}}},
\end{displaymath}
with $S_{i j}=0$ if users $i$ and $j$ haven't judged more than one item in common.
User $j$'s recommendation list is generated by stacking $v_{j \beta}$ in a vector, sorting the elements in descending order and following the procedure described in Sec.\ref{step:brank}. 
\subsection*{GRank}
GRank is a global ranking scheme. Objects are ranked according their popularity (number of votes): $k_{\alpha}$. Unlike B-Rank, CollabF and ZLZ-II, GRank takes not into account users personality, since it generates the same recommendation list for every user participating in the system. The ranking list is given by sorting the objects according their popularity in descending order.
\subsection*{Data sets}
Two data sets are used to B-Rank.
{\bf MovieLens} (movielens.umn.edu), a  web service from GroupLens (grouplens.org). Ratings are recorded on a five stars scale.
The data set contains $1682$ movies $\times$ $943$ users. Only $6,5\%$ of possible
votes are expressed.
{\bf Jester} (shadow.ieor.Berkeley.edu/humor), an online joke recommender system. The
data set contains $73421$ users $\times$ $100$ jokes. In contrast to MovieLens,
the data set from jester is dense: $75\%$ of all votes are expressed. The
rating scale are real numbers between $-10$ and $10$.

Apart from the sparsity and the dimensional ratio (number of object vs. number of users), the most fundamental difference
is the amount of a priori information accessible to users. People choose movies
they want to see on the basis of many different information sources. They know
actors, they read reviews, they ask friends for feedback etc. When users buy their tickets, they already did
a pre-selection. On the other hand no pre-selection is possible with online jokes.

In this sense, the two data sets are complementary. Tests on diverse data sets are more meaningful in general \cite{HeKo04}.
For a discussion on different performance aspects, see \cite{BlHuLa05}.
\subsection*{Performance evaluation}
 \label{metrics}
To test the algorithms the data are divided in two disjoint sets, a training set $S_{tr}$ and a test set $S_{ts}$. The training set is used to predict missing votes contained sin the test set.

Four different evaluation metrics were implemented: recall, precision, F1 and
diversity. The last is adopted from \cite{ZhKuLiMe09}.
Recall for user $i$ is defined as the number of recovered items $d_{i}$ in the top $N$ places of the recommendation
list, divided by the number of items $D_{i}$ in the test set for that user, thus $PR_{i}=d_{i}(N)/D_{i}$. Averaging over
all users gives the final score for recall $PR$.
Precision measures the number of recovered items in the top $N$ places divided by the length of the recommendation list $N$.
For user $i$ we have $PP_{i} = d_{i}(N)/N$. The overall precision $PP$ is obtained by averaging over all $PP_{i}$.

Increasing $N$ (length of the recommendation list)  usually increases recall and decreases precision at the same time. To balance
out these effects,  it is common to use the F1 metric, the geometrical mean of recall and precision:
$F1 = (2*PR*PP)/(PR + PP)$.

To test the diversity between different recommendation lists, $h(N)$ is used, a metric proposed in \cite{ZhKuLiMe09}.
The metric measures the diversity in the top $N$ places of two different recommendation lists.
$h_{ij}(N) = 1- (q_{ij}(N)/N)$, where $q_{ij}(N)$ denotes the number of common items in the top $N$ places
of list $i$ and $j$. $h_{ij}=1$ means there are no common items in the two lists, whereas $h_{ij}=0$ means complete match. Averaging over all
$h_{ij}(N)$ gives the population personalization level $h(N)$.

Each experiment was done on $20$ different instances - i.e. different splittings for training and test set with a fixed ratio (number of votes in the
test set vs. number of votes in the training set). Final scores for all metrics were obtained by averaging
over all instance results. All methods (B-Rank, ZLZ-II, GRank) were tested with  the same instances, to make a fair comparison.
\section{RESULTS}
\label{sect:results}
\subsection*{Numerical performance evaluation}
The main results of the numerical performance evaluation are collected in tables \ref{table:results1}-\ref{table:results8}. Bold figures indicate best result for a given evaluation
metric. The length of the recommendation list $N$ was set to $N=20$ and $N=10$ for all experiments. The performance improvement  is measured relative to the ZLZ-II algorithm.
There is a tendency toward higher improvements for shorter recommendation lists ($N=10$).
Best improvements are achieved for diversity between different recommendation lists.

\begin{table}[ht!]
{\small
\begin{tabular}{l l l l l  c}
&                B-Rank &                     ZLZ II &         CollabF &       GRank & Impr \\
PR     &  $\mathbf{0.34}$         & $0.30$       &$0.18$                   & $0.05$  &\\
PP  & $\mathbf{0.20}$        & $0.17$      &$0.15$                  & $0.04 $  &\\
F1        & $\mathbf{0.25}$            & $0.21$        &$0.14$                  & $0.04$  & $19\%$ \\
h  & $\mathbf{0.81}$         & $0.68$        &$0.70$                 & $0.19$  & $19\%$ \\
\end{tabular}
}
{\small
\caption{Movielens:  $N = 20$, test set = 20\% of expressed votes}}
\label{table:results1}
\end{table}

\begin{table}[ht!]
{\small
\begin{tabular}{l l l l l c}
&                B-Rank &                     ZLZ II &          CollabF &                       GRank & Impr\\
PR     &  $\mathbf{0.21}$        & $0.19$       &$0.11$                                         & $0.03$  &\\
PP  & $\mathbf{0.24}$        &$0.22 $       &$0.18$                                   & $0.04 $  &\\
F1        & $\mathbf{0.22}$            & $0.20$        &$0.12$                                  & $0.03$ & $10\%$\\
h  & $\mathbf{0.84}$         & $0.71$        &$0.74$                                   & $0.22$ & $18\%$ \\
\end{tabular}
}
{\small
\caption{Movielens:  $N = 10$, test set = 20\% of expressed votes}}
\label{table:results2}
\end{table}
\begin{table}[ht!]
{\small
\begin{tabular}{l l l l l c}
&                B-Rank & ZLZ II & CollabF & GRank & Impr\\
PR     &	$\mathbf{0.29}$        & $0.27$  &$0.12$          & $0.05$  &\\
PP  & 	$\mathbf{0.33}$        & $0.32$   &$0.32$        & $0.08$  &\\
F1         & 	$\mathbf{0.31}$        & $0.29 $   &$0.15$       & $0.06$  & $7\%$\\
h  & 	$\mathbf{0.77}$        & $0.64$    &$0.74$        & $0.15$  & $20\%$\\
\end{tabular}
}
{\small
\caption{Movielens for $N = 20$, test set = 70\% of expressed votes}}
\label{table:results3}
\end{table}
\begin{table}[ht!]
{\small
\begin{tabular}{l l l l l c}
&                B-Rank & ZLZ II &  CollabF & GRank & Impr\\
PR     &	$\mathbf{0.17}$          & $0.15$  &$0.07$             & $0.02$ &\\
PP  & 	$\mathbf{0.43}$         & $0.40$     &$0.37$         & $0.11$  &\\
F1         & 	$\mathbf{0.24}$          & $0.21 $    &$0.11$        & $0.03$ & $14\%$ \\
h  & 	$\mathbf{0.80}$         & $0.64$      &$0.78$        & $0.15$  & $25\%$ \\
\end{tabular}
}
{\small
\caption{Movielens for $N = 10$, test set = 70\% of expressed votes}}
\label{table:results4}
\end{table}
\begin{table}[ht!]
{\small
\begin{tabular}{l l l l l c}
&                		B-Rank & 				ZLZ II & 	CollabF&	GRank & 		Impr\\
PR     & 		$\mathbf{0.86}$    		& $0.72$    &$0.66$       & $0.48$ &	\\
PP & 		$\mathbf{0.38}$        	& $ 0.31$    &$0.33$       & $0.23$  &\\
F1         & 		$\mathbf{0.53}$       	& $0.43 $    &$0.41$ & $0.31$ &	$23\%$	\\
h  & 		$\mathbf{0.71}$		& $0.65$ 	&$0.66$ & $0.52$   &	$9\%$ \\
\end{tabular}
}
{\small
\caption{Jester $N = 20$, test set = 20\% of expressed votes}}
\label{table:results5}
\end{table}
\begin{table}[ht!]
{\small
\begin{tabular}{l l l l l c}
&                		B-Rank & 				ZLZ II & 	  CollabF&	GRank &		Impr\\
PR     & 		$\mathbf{0.72}$     	& $0.50$     &$0.49$      & $0.20$ 	&\\
PP & 		$\mathbf{0.59}$        	& $0.41$      &$0.46$     & $0.20$ 	&\\
F1         & 		$\mathbf{0.64}$       	& $0.45 $	&$0.44$ & $0.20$ 	 & $42\%$\\
h  & 		$\mathbf{0.80}$   		& $0.72$ 	&$0.75$ & $0.52$		& $11\%$ \\
\end{tabular}
}
{\small
\caption{Jester $N = 10$, test set = 20\% of expressed votes}}
\label{table:results6}
\end{table}
\begin{table}[ht!]
{\small
\begin{tabular}{l l l l l c}
&                		B-Rank & ZLZ II & CollabF & GRank &						Impr\\
PR     			& $\mathbf{0.66}$        	& $0.53$ &$0.44$          & $0.35$ 	&\\
PP  		& $\mathbf{0.72}$        	& $0.53$  &$0.69$        & $0.35$ 	&\\
F1         			& $\mathbf{0.68}$       	& $0.53$    &$0.51$       & $0.35$	& $28\%$ \\
h 			& $\mathbf{0.69}$	   	& $0.44$    &$0.52$       & $0.32$	& $57\%$ \\
\end{tabular}
}
{\small
\caption{Jester $N = 20$, test set = 70\% of expressed votes}}
\label{table:results7}
\end{table}
\begin{table}[ht!]
{\small
\begin{tabular}{l l l l l c}
&                		B-Rank & 				ZLZ II & 	CollabF&	GRank		&Impr\\
PR     & 		$\mathbf{0.49}$     	& $0.31$   &$0.25$        & $0.13$ 	&\\
PP & 		$\mathbf{0.90}$        	& $ 0.68$   &$0.75$        & $0.35$ 	&\\
F1         & 		$\mathbf{0.63}$       	& $0.42 $    &$0.36$	 & $0.19$ 	&$50\%$\\
h  & 		$\mathbf{0.80}$	       & $0.50$ 	&   &$0.63$   $0.33$		&$60\%$\\
\end{tabular}
}
{\small
\caption{Jester $N = 10$, test set = 70\% of expressed votes}}
\label{table:results8}
\end{table}

\subsection*{Computational issues \label{brankdet2}}
Two different computational aspects are investigated: 1) How to make an efficient 'real-time' recommendation without performing the matrix-vector multiplication needed by B-Rank and 2) An update algorithm for the transition matrix $P\rightarrow P_{new}$, avoiding matrix-matrix multiplication to calculate $P_{new}$. Note: the matrix-matrix multiplication is needed to calculate the adjacency matrix $A$ for the hypergraph.
Offline-Online tasks in B-Rank:
To calculate user $i$'s recommendation list $f_{(i)}$, one has to perform two matrix-vector multiplications - steps
$1$ and $2$ described in Sec.(\ref{step:brank}) - and an element-wise multiplication of two vectors. We can reduce
the effort to compute  the matrix-vector multiplications. The idea is simple: calculate object specific basis representations $b^{F}$ and $b^{B}$
for the forward and backward propagation vectors, independent of all users.
The recommendation task (online) for a user $i$ is then reduced to calculate a linear combination of the basis forward and backward propagation vectors.

{\bf Offline:}
The basis representations $b^{F}$ and $b^{B}$ vectors are defined as follows:
\begin{displaymath}
b^{F}_{\alpha} = P^{T}e_{\alpha}, \quad \quad  b^{B}_{\alpha} = Pe_{\alpha}.
\end{displaymath}
$e_{\alpha} \in \mathbb{R}^{O}$ is a natural basis vector, where the dimension $O$ is given by the number
of objects.

{\bf Online:}
The forward and backward propagation vectors for user $i$ are then given by:
\begin{displaymath}
f_{(i)}^{F} = \sum_{\alpha}c_{\alpha}b_{\alpha}^{F}, \quad \quad f_{(i)}^{B} = \sum_{\alpha}c_{\alpha}b_{\alpha}^{B}.
\end{displaymath}
With $c_{\alpha} = \chi_{(i)\alpha}$. The final calculation of user $i$'s ranking list $f_{(i)}$ is given by step $3$,
described in Sec.(\ref{subsect:brank}). Note: using this shortcut produces different figures in the recommendation lists, compared to the ones,  generated by the procedure in Sec.(\ref{subsect:brank}). However, the ranking (ordered list) will be the same.

The online part is easily done, since the calculation essentially reduces to calculate a linear combination of rows and columns from
the transition matrix $P$.

{\bf Update algorithm:}
the main effort to calculate the transition matrix $P$ consists of a matrix-matrix multiplication to compute the adjacency matrix $A$ of the hypergraph. A naive way to maintain the
system would be a re-calculation of $P \rightarrow P_{new}$, every time a user rated an object.
A simple update algorithm for $P \rightarrow P_{new}$ is given. The transition matrix $P$ is written like:
\begin{displaymath}
P(H)=\left[ G(H) \right]^{-1} F(H), \quad F(H) = HH^{T} - (xH^{T})^{D},
\end{displaymath}
\begin{displaymath}
G(H) = (y F(H))^{D}.
\end{displaymath}

The superscript $D$ denotes a diagonal matrix. $H$ is the incidence matrix defined in Sec.(\ref{step:brank}). $x$ and $y$ are row vectors
of appropriate format containing all ones. Then $xW$ is a vector containing the column sums of a matrix $W$.
The updated matrix $P_{new}(H)$ is defined as:
\begin{eqnarray*}
\label{equ:update}
P_{new}(H) & = &(P(H) + \Delta P(H)) \\ & = & [G(H) + \Delta G(H)]^{-1}[ F(H)+ \Delta F(H)].
\end{eqnarray*}
$\Delta X(H)$ denotes the change in $X$ when changing $H$.
For $\Delta G(H)$ and $\Delta F(H)$ we have:
\begin{displaymath}
\Delta G(H) = [y \Delta F(H)]^{D},
\end{displaymath}
\begin{eqnarray*}
\Delta F(H) &= &[(\Delta H)H^{T}] + [H(\Delta H^{T})] \\ &=& [(\Delta H)H^{T}] + [(\Delta H)H^{T}]^{T}.
\end{eqnarray*}

{\bf Single vote manipulation:} The update algorithm is investigated in more detail, in the case of one additional vote in the incidence matrix $H$.
To model a one vote change in $H$, the single-entry matrix $J^{ij} \in \mathbb{R}^{n \times m}$ is introduced, which is zero everywhere except in the $(i,j)$th entry, which is $1$.
Assume a matrix $A(n \times m)$ and a matrix $J^{ij}(m \times p)$, then
\bed
AJ^{ij} = [ 0 \quad  0 \cdots A_{i} \cdots 0 \quad 0]
\eed
is a $n \times p$ matrix with the i.th column of $A$ in place of the j.th column. Conversely, assume $A(n \times m)$ and $J^{ij}(p \times n)$.
Then, $J^{ij}A$ is a $p \times m$ matrix, with the j.th row of $A$ in the place of the i.th row. These operations are only column and row swapping of a matrix.

A single vote change is denoted as $\Delta H = J^{i,j}(O \times U)$. For $P_{new}(H)$ we have:
\begin{align}
\label{equ:upfinal}
P_{new}(H) &= \left[G(H) + y \left[ J^{ij}H^{T} + \left(J^{ij}H^{T}\right)^{T}\right]^{D}\right]^{-1}\\ & \quad \cdot [ F(H)+ J^{ij}H^{T} + \left(J^{ij}H^{T} \right)^{T}] \nonumber
\end{align}
Eq.(\ref{equ:upfinal}) is very efficient  - $O(o)$ at most instead of $O(u^{2}o)$.

{\bf Many vote manipulation:} the generalization of single vote manipulations is straightforward,
since a many vote update is represented by a combination of single vote updates.
\section{DISCUSSION}
\label{sect:disc}
Results show significant performance improvement in all experiments. B-Rank is able to perform well on complementary data sets. However, like every experiment with recommender systems, results are always 'bound' on used data sets. There is no guaranty to obtain similar results for different data. 

The best improvement, compared to ZLZ-II, is achieved for inter list diversity. This result highlights the fact, that B-Rank can cope with users personality. From real world experiments we know, that higher diversity is positive correlated to user satisfaction in general \cite{NeRiKo06}. However, user satisfaction is hard to measure in off-line experiments and user feedback is needed to draw robust conclusions.    

An extension to the ZLZ-II algorithm was proposed by \cite{LiZh09}, where the authors reached a comparable performance for diversity like B-Rank in the movielens dataset. Their method includes a tuning parameter $\lambda$. B-Rank in contrast is parameter free and  therefore easier to implement and maintain.

Extensions to B-Rank may increase improvements again. One extension to the presented basic B-Rank algorithm is a non constant weight matrix $W$. This will be discussed in a follow up paper. Another extension is to take into account $n$-step propagation (indirect connections between two objects $\alpha$ and $\beta$). Tests for different $n>1$  significantly dropped recall, precision and
inter list diversity performance as well. One explanation for this behavior is a propagation reinforcement of popular items. 

The basic version of B-Rank can be extended by introducing a user dependent parameter $\eta(i)$, controlling the contribution of backward and forward propagation: $f_{(i)} = \left(f_{(i)}^F \right)^{\eta(i)}$ \# $\left(f_{(i)}^B \right)^{1-\eta(i)}$. Such a parameter is a fine tuning of user $i$'s preferences for popular and niche objects. Also a user independent, global $\eta$ is possible. All these extensions increase computational compexity, since the system have to learn the 'correct' parameters.   

Extensions and non trivial weight matrices $W$ will be investigated and presented in a follow up paper.
\section{SUMMARY}
\label{sect:sum}
In this paper B-Rank, a new top $N$ recommendation algorithm, is proposed. The algorithm is based on a random walk model on hypergraphs. B-Rank is easy to implement and needs no parameter tuning. The algorithm outperforms other state-of-the-art methods like ZLZ-II \cite{ZhLuZh09} and Collaborative Filtering in terms of accuracy and inter list diversity.

B-Rank is able to find interesting 'blockbusters' and niche objects as well.  The algorithm is very promising for different applications, since it produces good results for sparse and dense voting matrices as well.
Furthermore, a simple recommendation list update algorithm is introduced, which dramatically reduces computational complexity, Sec.\ref{sect:addmat}.
\section{ADDITIONAL MATERIAL}
\label{sect:addmat}
\begin{figure}[ht!]
\centerline{\includegraphics[width=0.2\textwidth]{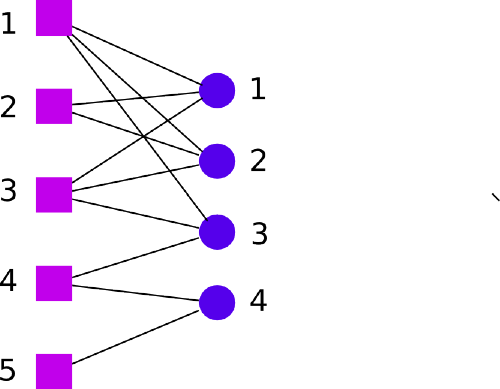}}
\caption{Toy net to illustrate B-Rank. Circles represent hyperedges (users), squares are hypervertices, i.e. objects.
The votes are illustrated as links between objects and users.
}
\label{app:toynet}
\end{figure}

To highlight various aspects of B-Rank, a toy network Fig.(\ref{app:toynet}) is introduced. For simplicity all links between objects and users are equally weighted
$w_{i}=1$ $\forall i$.

First, some general aspects are discussed, second it is shown, how all aspects are well captured by the B-Rank algorithm.

{\bf Case A: huge audience in common.}
Intuitively, two objects $\alpha$ and $\beta$ are similar to each other, when they share many users - i.e. they have many hyperedges in common. Let's assume object $\alpha$ and object $\beta$ share many users and user $i$ voted for $\alpha$ but didn't vote for $\beta$ yet. Then it's reasonable to recommend $\beta$ to user $i$. Such a recommendation strategy clearly favors ``blockbusters'', objects rated by almost every user in the community (e.g objects $1$ and $3$ in the toy network Fig.(\ref{app:toynet}).

{\bf Case B: exclusive audience.}
Look at object $5$ in the toy network: this object is exclusively rated by user $4$. Moreover, object $4$ and object $5$ share only user $4$ and object $4$ was
not rated by many other users. In this sense, object $4$ and $5$ have an exclusive audience in common. It is reasonable to mark these objects as very similar and to recommend one of them to users who have not rated both.

{\bf Do the random walk.}
Aa path $(\alpha \rightarrow \beta)$ is defined as an ordered triple $\{\alpha ,i, \beta\}$ with $\alpha \ne \beta$ (i.e object, user, object). The transition probability $P_{\alpha \beta}$ in Eq.(\ref{equ:brank}) counts the number
of paths (triples) starting at $\alpha$ and ending at $\beta$, divided by the number of all paths starting at $\alpha$.
Examples: for $P_{12}$ we count $6$ paths starting at object $1$. Two of them ending at object $2$, thus $P_{12}=1/3$. For
$P_{13}$ we count again $6$ paths starting at $1$, and three paths ending at object $3$, thus $P_{13}=1/2$. Note, that $P_{\alpha \beta} \ne P_{\beta \alpha}$ in general, and $P_{\alpha \alpha} = 0$ $\forall \alpha$.

{\bf Put everything together.}
To demonstrate the effect of forward and backward propagation in B-Rank  we use a basic
preference vector $\chi = \left[0 ,0,0,1,0 \right]^{T}$ and the topology of the toy net in Fig.(\ref{app:toynet}). For the forward propagation  $f^{F} = P^{T} \chi$ we get:
\begin{displaymath}
f^{F} = [1/3, 0, 1/3, 0, 1/3]^{T}
\end{displaymath}
The obtained figures for objects $\alpha \ne 4$ indicate the probability for a random walker starting at object $4$ and landing at $\alpha \ne 4$. Note, the scores  are the same for objects $1,3,5$. Object $2$ obtains no score, because there is no simple path
from object $4$ to object $2$. Object $4$ obtains no score since the path $\{4,i,4\}$ is not a valid path per definition. 
For the backward propagation  $f^{B} = P \chi$ we get:
\begin{displaymath}
f^{B} = [1/6, 0, 1/6, 0, 1]^{T}
\end{displaymath}
The backward propagation vector contains the probabilities for a random walker starting at objects $\alpha \ne 4$ and landing at object $4$.
We observe the same score for object $1$ and object $3$, but a much higher score for object
$5$, since the probability for a random walker starting at object $5$ and ending at object $4$ is
much higher, then the probability reaching object $4$ from another node.

The final score $f$ is given by the element wise multiplication of $f^{F}\#f^{B}$. Thus
\begin{displaymath}
f = [1/18, 0, 1/18, 0,1/3]^{T}
\end{displaymath}
The final score for each object $\alpha \ne 4$ has a simple interpretation: it is the probability for a random walker starting at object $4$, visiting object $\alpha$ and come back to object $4$.

The higher score of object $5$ makes sense in the given setup,
because objects $4$ and $5$ share an exclusive audience, furthermore object $4$ is only
'loosely' connected to all  other objects.

B-Rank captures well the possible configurations described in case A and B. If an object $\alpha$ has many links and shares most of them with another object $\beta$,
then $\beta$ is reached with higher probability then other objects, less connected (number of paths) to $\alpha$. On the other
hand, if an object $\alpha$ has many connections, but shares exclusively some hyperedges (users) with an object $\beta$, then $f^{F}$ may give low resource to $\beta$,
but $f^{B}$ will give a high score to the same object  $\beta$. In summary: B-Rank takes into account propagation of popular and niche objects as well.

Introducing hyperedge weights, described in Sec.(\ref{subsect:brank}), is a generalization of the procedure described in this appendix. It is not clear, what weight function is an appropriate choice. This issue will be investigated in an follow up paper.
\section*{ACKNOWLEDGMENT}
I thank Matus Medo and the Econophysics team of the University of Fribourg in Switzerland for numerous
discussions and valuable comments.

\bibliography{refer}
\bibliographystyle{IEEEtran}

\end{document}